\documentclass[12pt]{iopart}

\usepackage{iopams} 
\usepackage[english]{babel}
\usepackage{graphicx}
\usepackage{bm}
\usepackage{latexsym}
\usepackage{mathrsfs}
\usepackage{pifont}
\usepackage{pstricks,pst-node,pst-text,pst-3d}
\usepackage[T1]{fontenc}
\usepackage{ulem}  
\usepackage{caption} 
\newcommand{\fig}[1]{figure~\ref{#1}}
\newcommand{\Fig}[1]{Figure~\ref{#1}}
\newcommand{\tab}[1]{Table~\ref{#1}}
\newcommand{\eq}[1]{(\ref{#1})}

\usepackage{ulem}  
\usepackage{color}

\newcommand{\icspace}{IC-space}
\newcommand{\qA}{q_{\rm A}}
\newcommand{\qB}{q_{\rm B}}
\newcommand{\rate}{w}

\newcommand{\pole}{arm}

\newcommand{\GrotsymB}{directional rotatory motion} 
\newcommand{\adticMotAsym}{directionality of the rotatory motion} 
\newcommand{\eqSym}{invariance} 
\newcommand{\asymMeas}{directionality measure} 
\newcommand{\antisymticPS}{anti-symmetric} 

\newcommand{\symRot}{symmetric rotor} 
\newcommand{\asymRot}{asymmetric rotor} 
\newcommand{\stochJ}{hopping-process}

\begin{document}

\title{Rotational directionality via symmetry-breaking in an electrostatic motor}

\author{A Celestino$^1$, A Croy$^1$, M W Beims$^{2,1}$, A Eisfeld$^1$}
\address{$^1$ Max Planck Institute for the Physics of Complex Systems,  N\"othnitzer Str.~38, 01187 Dresden, Germany}
\address{$^2$ Departamento de F\'\i sica, Universidade Federal do Paran\'a,  81531-980 Curitiba,  Brazil}
\ead{eisfeld@pks.mpg.de}
\vspace{10pt}
\begin{indented}
\item[]\today
\end{indented}
%
\begin{abstract}
We theoretically investigate how one can achieve a preferred rotational direction for the case of a simple electrostatic motor.
The motor is composed by a rotor and two electronic reservoirs. Electronic islands on the rotor can exchange electrons with the
reservoirs. An electrostatic field exerts a force on the occupied islands. The charge dynamics and the electrostatic field drive
rotations of the rotor. Coupling to an environment lead to damping on the rotational degree of freedom.
We use two different approaches to the charge dynamics in the electronic islands: hopping process and mean-field.
The hopping process approach takes into account charge fluctuations, which can appear along Coulomb blockade effects in
nanoscale systems. The mean-field approach neglects the charge fluctuations on the islands, which is typically suitable for larger systems.
We show that for a system described by the mean-field equations one can in principle prepare initial conditions to obtain a desired rotational
direction. In contrast, this is not possible in the stochastic description.
However, for both cases one can achieve rotational directionality by changing the geometry of the rotor.
By scanning the space formed by the relevant geometric parameters we find optimal geometries, while fixing the dissipation and driving parameters.
Remarkably, in the hopping process approach perfect rotational directionality is possible for a large range of geometries. 
\end{abstract}
%


%
\vspace{2pc}

\submitto{\NJP}
%
%
\maketitle
%
%

\section{Introduction}

\begin{figure}[tb]
  \centering
  \includegraphics*[width=1.0\columnwidth]{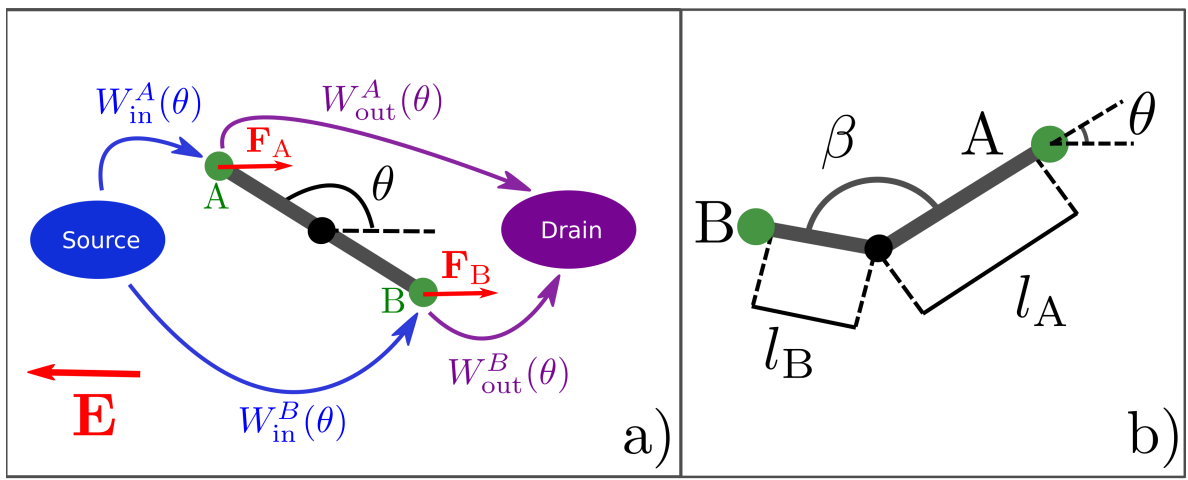}
  \caption{Sketch of the setup and its relevant parameters.  The setup is composed by a rotor (shaft, arms and islands A and B) and two negative charge reservoirs (source and drain). The rotor's orientation is given by $\theta$. The islands A and B are (dis)charged by the source with rate $W_{\rm in}^{\rm A/B}$($W_{\rm out}^{\rm A/B}$). An electrostatic field $\vec{E}$ exerts a force $\vec{F}$ on the charged islands. While in a) the {\it symmetric} rotor is shown, in b) the {\it asymmetric} rotor is displayed together with the parameters to describe the deviation from the symmetric case. }
  \label{drawing}
\end{figure}

Rotary motors are paradigmatic devices which transform input energies in controlled rotations. These rotations can be employed to yield work, powering a nanomachine for instance \cite{Article,Motors2007}. In the micro-scale these devices have been assembled with nanowires \cite{Kim2015548,Guo2015} and microfabricated gears \cite{Maggi2015}. Since the advent of the first synthetic molecular motor \cite{koumura1999light,Kelly1999}, many different designs for nano-scale motors were proposed \cite{Bailey2008,Bustos-Marun2013,Morin2006,croy2012dynamics,wang2008nanoscale,smirnov2009bio} and realized \cite{leigh2003unidirectional,Fennimore2003,van2005unidirectional,Perera2012,kottas2005artificial,Guix2014nano,Murphy2014,
Mishra2015}. Significant steps towards application of rotary motors have been made, for instance they have been used to propel a molecule in a Cu(111) surface \cite{Kudernac2011}, rotate microscale objecs \cite{eelkema2006} and control chemical reactions \cite{Pons2011}. These devices can be driven in many different ways, using e.g.\ light \cite{koumura1999light,van2005unidirectional,Maggi2015}, chemical energy \cite{Kelly1999,Fletcher2005}, electronic current \cite{Perera2012,Murphy2014,Mishra2015} and electric fields \cite{Fennimore2003,Kim2015548}. Recent publications \cite{croy2012dynamics,wang2008nanoscale,smirnov2009bio} consider single-electron tunnelling to drive a nanoelectromechanical rotor.
In the present work, we will focus on the setup of Ref.~\cite{croy2012dynamics}. In this setup, sketched in \fig{drawing}, a mechanical rotor is placed between
two leads at different chemical potentials. The rotor has islands (capacitors) attached to its tips and electrons can tunnel from
the ``source'' lead to the islands and from the islands to the ``drain'' lead. An electrostatic field between the leads induces a
force on the charged islands, which eventually leads to a torque on the rotor.

The charging and discharging processes depend fundamentally on the size of the islands, and consequently on the size of the device.
For nano-scale devices where the Coloumb blockade plays a role, the islands can only host a small number of electrons (in the extreme case only one). In this case, charge fluctuations caused by the stochastic nature of the tunnelling process lead to torque fluctuations on the rotor. On the other hand, for larger devices where Coulomb blockade is not relevant, charge fluctuations are irrelevant, since they are very small compared to the mean charge on the islands. In Ref.~\cite{croy2012dynamics}  it has been discussed that a mean-field description, which is valid for macroscopic rotors, is also able to reproduce certain aspects of the stochastic description which describes rotors in the Coulomb blockade regime. In particular, it has been found that by increasing the ratio of the driving and damping parameters one obtains a regime where the rotor is performing rotational motion and the (averaged) angular momentum is the same in the stochastic and the mean field description.

For many applications it is crucial to attain control over the directionality of the rotations of the motor. In the mean-field description, the rotational direction depends on the specific initial condition (IC). The IC is characterized by the initial angular momentum $L(t=0)=L_0$, the initial angle $\theta(t=0)=\theta_0$, and the initial electronic charges $Q_{\rm A}(t=0)$ and $Q_{\rm B}(t=0)$ on the islands. However, the experimental preparation of a
specific IC can be difficult, imprecise or even inconvenient, depending on the system scale, suppression of noise-sources, etc.
Moreover, imperfections in the rotor fabrication, namely in the length of its arms, angle between them, capacities of the islands, etc. could also lead to uncertainty over the rotational direction. In the stochastic description the rotation direction can either vary from realization to realization independently of the IC (weak electric field, strong damping), or even alternate in time within a single trajectory (strong electric field, weak damping). 

Another way of attaining rotational directionality is to make a rotation direction ``preferable''. With ``preferred direction'' we mean that most of, or even all ICs(realizations) lead to rotations in this direction. For a
perfectly \symRot{}, it is intuitively clear that no directionality is achieved: for each IC(realization) leading to a certain rotation direction there exists another one leading to the opposite direction. This gives rise to the question which symmetries of the rotor one has to break in order to achieve \adticMotAsym{}.

In this paper, we show that by changing simultaneously the angle between the \pole{}s of the rotor and their length, it is possible to introduce a preferred direction of rotation. 
We search in particular for combinations of the parameters which lead to \GrotsymB{}. 
We will also see that it is not sufficient to change either the angle or the length alone. The parameter space of the studied system presents regions where all ICs end up rotating in the same direction and with the same $L(t)$. It means that no control over IC is required to achieve a certain direction of rotation, and that within some constraints, the rotor fabrication can also present imperfections.

The paper is organized as follows. Section \ref{sec:model} presents
the model for the electrostatic motor, which includes the mean-field equations of 
motion, and discussions about relevant quantities. 
While Sec.~\ref{sec:symmetric} presents some features of the 
\symRot{}, Sec.~\ref{sec:asymmetric} shows how \GrotsymB{} can be achieved in the mean-field approach. In Sec.~\ref{sec:stochastic} we treat the case of nanoscale 
devices using a \stochJ{} approach for the tunnelling and in Sec.~\ref{sec:conclusion} we summarize our 
results.

\section{Mean-field equations and definitions}
\label{sec:model}
In this section we  present mean-field equations of motion for the rotor, discuss as special example the \symRot{}, and provide definitions of several quantities which will be of interest in the following sections.

\subsection{Mean-field equations of motion}
We consider a rotor composed of insulating arms and electronic islands A and B placed at their ends as sketched in \fig{drawing}. 
The rotor is immersed in a region with a homogeneous, time-independent electric field $\mathbf{E}$, directed along the horizontal axis.
To specify the orientation of the rotor, we use the angle $\theta$ between the rotor arm connected to A and the horizontal line as indicated in \fig{drawing}.
We allow the two arms of the rotor to be tilted with respect to each other, and denote the angle between them  by $\beta$.
The islands can host charges $Q_{\rm A}$ and $Q_{\rm B}$ up to a certain limit $Q_{\rm max}$ (chosen to be equal for both islands) which is given by their capacitances $C_{\rm A}=C_{\rm B}$. 
As in Ref.~\cite{croy2012dynamics}, we start by considering the dynamics using a mean-field approach where charge fluctuations are neglected and the charge dynamics is purely deterministic. This mean-field allows one already to understand the fundamentals of the symmetry-breaking and is strictly valid for macroscopic devices (see Appendix \ref{sec:rotor_setup_class}).The \stochJ{} for the tunneling, which is more suitable to nanoelectromechanical devices, will be considered in Section \ref{sec:stochastic}.
The equations that describe the charge dynamics on the islands in the mean-field the are given by
\numparts
\begin{eqnarray}
    \frac{{\rm d}Q_{\rm A}}{{\rm d}t} &= W_{\rm in}(\theta)(Q_{\rm max}-Q_{\rm A}) - W_{\rm out}(\theta)Q_{\rm A}, \label{eq:dtQ1} \\
    \frac{{\rm d}Q_{\rm B}}{{\rm d}t} &= W_{\rm in}(\theta+\beta)(Q_{\rm max}-Q_{\rm B}) - W_{\rm out}(\theta+\beta)Q_{\rm B} \label{eq:dtQ2},
\end{eqnarray}
\endnumparts
where $t$ is time, $W_{\rm in}(\theta')$ and $W_{\rm out}(\theta')$ are charging and discharging rates, respectively, and their argument $\theta'=\theta$ for the island A and $\theta'=\theta+\beta$ for the island B. 
We assume $W_{\rm in/out}(\theta)$ to be positive  even functions (their functional dependence on the angles $\theta$ and $\theta+\beta$ is going to be specified in subsection \ref{specW}). 
In the following we use dimensionless quantities. We define $\tau= t/t_0$ ($t_0$ is a characteristic time-scale for the electronic dynamics), $\qA(t)=Q_{\rm A}(t)/Q_{\rm max}$ and $\qB(t)=Q_{\rm B}(t)/Q_{\rm max}$ and  $\rate_{\rm in/out}(\theta')=t_0W_{\rm in/out}(\theta')$.
Then the dimensionless version of equations \eq{eq:dtQ1} and \eq{eq:dtQ2} is
\numparts
\begin{eqnarray}
    \frac{d\qA}{d\tau} &= \rate_{\rm in}(\theta)(1-\qA) - \rate_{\rm out}(\theta)\qA, \label{eq:dtq1} \\
    \frac{d\qB}{d\tau} &=\rate_{\rm in}(\theta+\beta)(1-\qB) - \rate_{\rm out}(\theta+\beta)\qB \label{eq:dtq2}.
\end{eqnarray}
\endnumparts
To realize these mean-field equations experimentally one can use for example RC circuits, where the islands are capacitors. Their capacitances have an angle dependence $C_{\rm A}(\theta)=[R(W_{\rm in}(\theta)+W_{\rm out}(\theta))]^{-1}$ and $C_{\rm B}(\theta) = [R(W_{\rm in}(\theta+\beta)+W_{\rm out}(\theta+\beta))]^{-1}$, where $R$ is the resistance (details are given in appendix \ref{sec:rotor_setup_class}). 

Since the rotor is placed within an electric field, the islands experience a force when they are charged.
Within the mean-field approach one finds the following set of coupled equations for the dynamics of the mechanical degrees of freedom (For a derivation we refer to appendix \ref{derivAng})
\numparts
\begin{eqnarray}
    \frac{\partial}{\partial \tau}\Lambda &=-\eta_0\sqrt{2}\left(\sin{\chi}\sin{(\theta)}\qA+\cos{\chi}\sin{(\theta+\beta)}\qB\right)-\gamma\Lambda ,\label{total1} \\
    \frac{\partial}{\partial \tau}\theta &= \Lambda . \label{total2}
\end{eqnarray}
\endnumparts
Here, we have introduced the dimensionless angular momentum 
$\Lambda=t_0 L/I$, where $L$ is the actual angular momentum and $I$ is the moment of inertia of the rotor.
The asymmetry parameter $\chi\in[0,\pi/2]$ is given by $\tan{\chi}=l_{\rm A}/l_{\rm B}$, with $l_{\rm A(B)}$ being the length of the arm containing island A(B). The driving strength $\eta_0>0$ is
\begin{equation}
\eta_0=\sqrt{\frac{l_{\rm A}^2+l_{\rm B}^2}{2}}\left[\frac{t_0^2 |Q_{\rm max}|E}{I}\right],
\end{equation}
with $E=|\vec{E}|$ being the electric field strength.
A surrounding environment leads to the phenomenological damping-term,  whose strength is given by the dimensionless viscosity parameter $\gamma$. The dimensionless quantities can be seen in \tab{dlqtts}.

Note that $\qA(t)$ and $\qB(t)$ are determined by equations \eq{eq:dtq1} and \eq{eq:dtq2}.
Since the right hand side of equations \eq{eq:dtq1} and \eq{eq:dtq2} depends on the angle $\theta$ we have to solve the non-linear system of coupled equations \eq{eq:dtq1}-\eq{total2}.

\begin{table}
\centering
\captionsetup{justification=centering}
\caption{\label{dlqtts}Dimensionless quantities and their definitions.}
   \begin{tabular} {| c | c | c |}
	\hline
	\multicolumn{1}{| c}{Quantity} & \multicolumn{1}{c}{Symbol} & \multicolumn{1}{c |}{Formula} \\
	\hline
	Time & $\tau$ & $\frac{t}{t_0}$ \\
	\hline
	Charge & $q_{\rm A/B}$ & $\frac{Q_{\rm A/B}}{Q_{\rm max}}$ \\
	\hline
	Charging rate & $w_{\rm in/out}(\theta)$ & $t_0W_{\rm in/out}(\theta)$ \\
	\hline
	Angular momentum & $\Lambda$ & $\frac{t_0L}{I}$ \\
	\hline
	Asymmetry parameter & $\chi$ & $\arctan{\left(\frac{l_{\rm A}}{l_{\rm B}}\right)}$ \\
	\hline
	Driving strength & $\eta_0$ & $\sqrt{\frac{l_{\rm A}^2+l_{\rm B}^2}{2}}\left[\frac{t_0^2 |Q_{\rm max}|E}{I}\right]$ \\
	\hline
\end{tabular}
\end{table}

\subsection{Quantities of interest}\label{quantities}
We are interested in the influence of initial conditions on the steady-state behavior of the system.
In particular we are interested in the time-average of the angular momentum (sign and  magnitude) and charge current.
The time-averaged angular momentum is defined as
\begin{equation}
    \langle\Lambda\rangle=\frac{1}{\mathcal{T}-\tau_{\rm tran}}\int_{\tau_{\rm tran}}^{\mathcal{T}}\Lambda (\tau) d\tau,
    \label{time-averaged}
\end{equation}
where $\tau_{\rm tran}$  is a typical transient time \footnote{In deterministic systems it is well defined that a dissipative system needs a certain time $\tau_{\rm tran}$ [depending on the initial conditions] to reach with a certain precision an asymptotic manifold in phase space called attractor \cite{Lichtenberg1992}. For 
systems described by stochastic differential equations one still has the concept of transient-time, although the concept of attractors as
asymptotic manifolds in phase space becomes meaningless \cite{Lai2011}.} and $\mathcal{T}$ is sufficiently large  to make $\langle\Lambda\rangle$ reach its stationary value.

We also investigate the time-averaged electronic current that flows through the system, and its dependence on the ICs.
The dimensionless time-averaged current is
defined as
\begin{eqnarray}
\langle J\rangle =\frac{1}{\mathcal{T}-\tau_{\rm tran}}\int_{\tau_{\rm tran}}^{\mathcal{T}}[\rate_{\rm out}(\theta(\tau))\qA(\tau)+ \rate_{\rm out}(\theta(\tau)+\beta)\qB(\tau)]d\tau,
\label{Jexp}
\end{eqnarray}
and the physical current is then given by $\langle J\rangle Q_{\rm max}/t_0$. Again, 
we consider the same transient-time $\tau_{\rm tran}$ as in equation \eq{time-averaged} and $\mathcal{T}$ to be large enough such that 
$\langle J\rangle$ reaches its stationary value after $\mathcal{T}$. 
In order to calculate $\langle\Lambda\rangle$ and $\langle J\rangle$ we solve the set of equations \eq{eq:dtq1}-\eq{total2} numerically, and use the trajectories to calculate the time-integrals \eq{time-averaged} and \eq{Jexp}.

\subsection{Specific form of the (dis)charging rates}
\label{specW}
Although our discussion does not require any specific form of $\rate_{\rm in/out}(\theta')$ we will use as an example the functional form adopted in Ref.~\cite{croy2012dynamics} to describe a nano-scale electromechanical rotor, namely
\numparts
\begin{eqnarray}
    \rate_{\rm in}(\theta') &= {\rm e}^{-\xi\cos(\theta')}, \label{rates1} \\
    \rate_{\rm out}(\theta') &={\rm e}^{\xi\cos(\theta')}. \label{rates2}
\end{eqnarray}
\endnumparts
In Ref.~\cite{croy2012dynamics} this particular form of $\rate_{\rm in/out}(\theta')$ stems from the exponentially decreasing (with respect to the lead-island distance) tunneling rates \footnote{In the nanoscale realization the dimensionless parameter $\xi$ is related to the so-called tunneling length $\lambda$ by $\xi=l/\lambda$, where $l=l_{\rm A,B}$ is the length of the \pole{} attached to A or B \cite{croy2012dynamics}, as one can identify in \fig{drawing}. The tunneling length is determined mainly by the choice of materials used to build the leads as well as the systems implemented as islands implementation.}. 
In this work we will take $\xi$ to be identical for the two islands and for source and drain. \footnote{Since $l=l_{\rm A,B}$ is the length of the rotor's \pole{}s, one sees immediately that one needs specific tunneling lengths $\lambda_{A,B}$ in order to keep $\xi$ independent on the considered island. This could be done, for instance, using different materials for the islands.}

\section{Symmetric rotor}
\label{sec:symmetric}

For reference, we start with the dynamics of the \symRot{} ($\beta=\pi$ and
$\chi=\pi/4$), which has aligned arms with the same length ($l_{\rm A}=l_{\rm B}$).
From now on, we will consider the charge IC $q_{\rm A}(\tau\!=\!0)=q_{\rm B}(\tau\!=\!0)=0$.

\subsection{Invariances of the \symRot{} }\label{symSymrot}
In the case of the \symRot{}, solutions of the coupled equations \eq{eq:dtq1}-\eq{total2} possess the following invariance: Substituting $\tilde{\Lambda}=-\Lambda$ and $\tilde{\theta}=-\theta$ {\it simultaneously} in equations \eq{eq:dtq1}-\eq{total2}, i.e.\ changing the direction of rotation and reflecting around $\theta=0$, one obtains the same equation. This invariance (with the rotation inversion) occurs also for reflections around $\theta=\pi/2$ and around $\theta=3\pi/2$. Due to these invariances, for every counter-clockwise (positive $\Lambda$) rotating solution with initial condition $(\Lambda_0,\theta_0)$ there is another clockwise (negative $\Lambda$) rotating one with initial condition $(-\Lambda_0,-\theta_0)$. 
This \eqSym{} will be of fundamental importance when we consider the time-averaged angular momentum $\langle\Lambda\rangle$, which is a function of the IC, averaged over $\theta_0$ for a rotor initially discharged ($\qA(\tau\!=\!0)=\qB(\tau\!=\!0)=0$) and at rest ($\Lambda_0=0$)
\begin{equation}
\overline{\langle\Lambda\rangle}=\frac{1}{2\pi}\int_{0}^{2\pi}\!\!\!{\rm d}\theta_0\langle\Lambda\rangle,
\label{overDef}
\end{equation}
which becomes exactly zero in the presence of this \eqSym{}.
\subsection{\icspace{} of the \symRot{}}

It turns out that the steady-state dynamics depends sensitively on the ratio $\eta_0/\gamma$ \cite{croy2012dynamics}.
For a given $\xi$ one has three dynamical regimes: for very small $\eta_0/\gamma$ there is no motion at all and the rotor rests in the position $\theta=\pi/2$. For larger $\eta_0/\gamma$ there is an interval  where oscillations are found. Above a certain threshold of $\eta_0/\gamma$ rotational motion sets in where the angular momentum increases approximately as the square root of $\eta_0/\gamma$. As said before, here we focus on parameters where rotational motion occurs and we will consider the case $\xi=2$ as an example.

\begin{figure}[tb]
  \centering
  \includegraphics*[width=0.49\columnwidth]{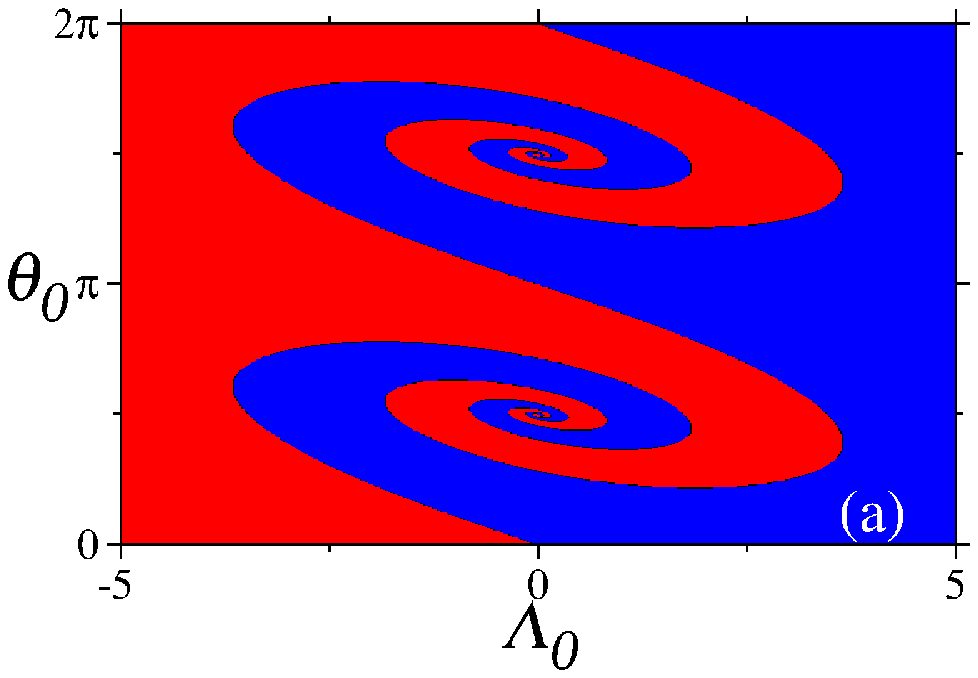}
  \includegraphics*[width=0.49\columnwidth]{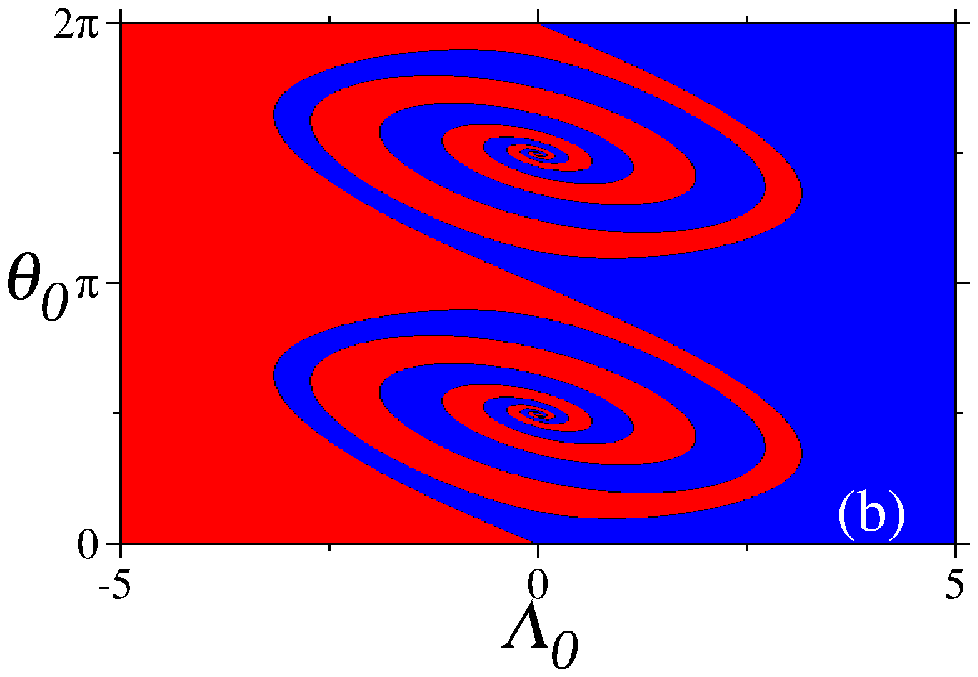}
 \caption{Time averaged angular momentum (color) $\langle \Lambda \rangle$ as a function of the
 ICs $(\Lambda_{\rm 0},\Theta_{\rm 0})$ for $\qA(\tau\!=\!0)=\qB(\tau\!=\!0)=0$, $\xi=2.0, \gamma=1$ and (a) $\eta_0/\gamma=10.0$ 
 (b) $\eta_0/\gamma=5.0$. Red stands for clockwise (negative $\Lambda$) and blue for counter-clockwise (positive $\Lambda$) rotation direction.}
  \label{220613fig1ab}
\end{figure}

Figure \ref{220613fig1ab} compares the behavior for two values of $\eta_0/\gamma$. Red stands for clockwise (negative $\Lambda$) and blue for counter-clockwise (positive $\Lambda$) rotation direction. Throughout the \icspace{}, for both negative and positive $\langle\Lambda\rangle$, one has the same $|\langle\Lambda\rangle|$.
In both plots one observes intertwined patterns spiraling towards $(\theta_0=\pi /2,\Lambda_0=0)$ and $(\theta_0=3\pi /2,\Lambda_0=0)$. These points correspond to the standstill situation ($\Lambda=0$) with the rotor in a vertical orientation. 
Both locations are fixed points in phase space: Once the system reaches those states, it stays there forever. The spiraling pattern around these fixed points can be understood as follows: the rotor transiently oscillates and as it swings it gains
energy until it reaches the rotatory steady-state. In each half oscillation
the rotor has the chance to set into the rotatory steady-state with a certain rotation direction. If it does not have enough energy, it performs another half oscillation gaining more energy and having the opportunity to start rotating into the opposite direction. This process is repeated until the stationary rotatory state is reached, whose direction depends on the IC of the rotor. 
For ICs closer to the fixed points a larger number of oscillations is required to reach the steady-state
and small changes of the IC can lead to a different direction of rotation. 
This fact causes the complex spiraling pattern seen in the figures. 
As one decreases the
driving strength $\eta_0$ the rotor gains energy more slowly and therefore more and more oscillations are required in order to reach the rotatory regime, which in turn leads also to finer spiraling patterns, as one sees in \fig{220613fig1ab} (b).

In \fig{220613fig1ab} one can also see a consequence of the invariance discussed above in subsection \ref{symSymrot}, namely an anti-symmetry around $\theta_0=0$ ($\theta_0=\pi$) and $\theta_0=\pi/2$($\theta_0=3\pi/2$) and $\Lambda_0=0$. 
This anti-symmetry implies in particular that $\overline{\langle\Lambda\rangle}=0$.
In the following section we will show how one can induce a preferred rotation direction, i.e.\ to make $\overline{\langle\Lambda\rangle}\neq 0$, even in the case when the angular momentum is initially zero.

Now we turn our attention to the second quantity of interest defined in subsection \ref{quantities}, namely the time-averaged current $\langle J\rangle$.
The time-averaged current only depends on the magnitude (and not on the sign) of the angular momentum.
Therefore all ICs lead to the same time-averaged current and we do not present a plot of $\langle J\rangle$ as a function of the ICs 
$(\Lambda_0,\theta_0)$.

\section{\asymRot{}}
\label{sec:asymmetric}
In this section we investigate how one can obtain $\overline{\langle\Lambda\rangle}\neq 0$ by breaking the rotor's symmetry.

\subsection{Changing either $\beta$ or $\chi$: No \adticMotAsym{}}

\begin{figure}[tb]
  \centering
  \includegraphics*[width=0.49\columnwidth]{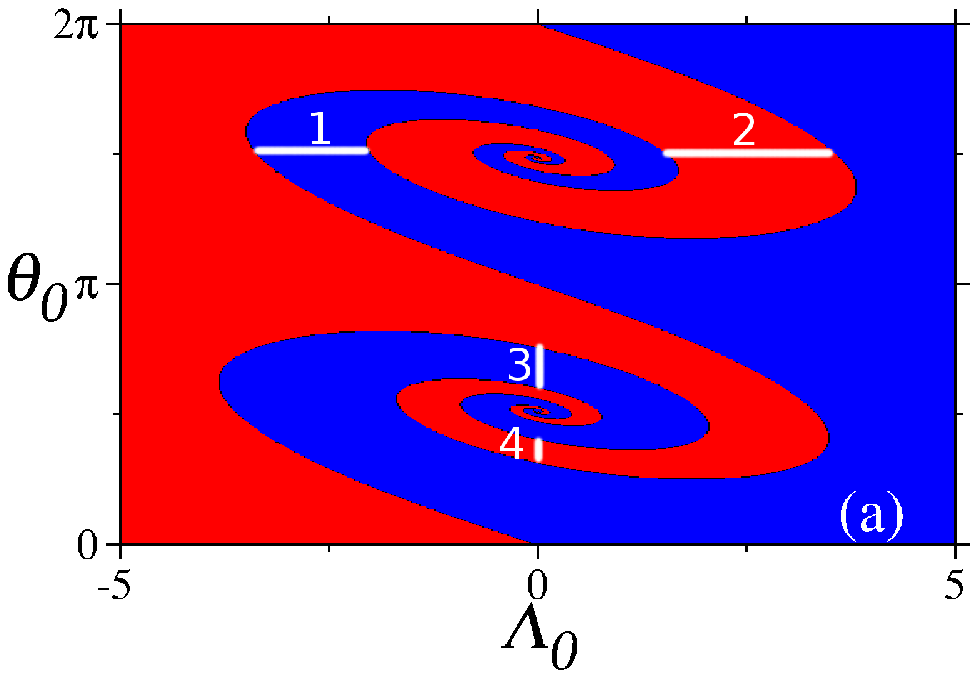}
  \includegraphics*[width=0.49\columnwidth]{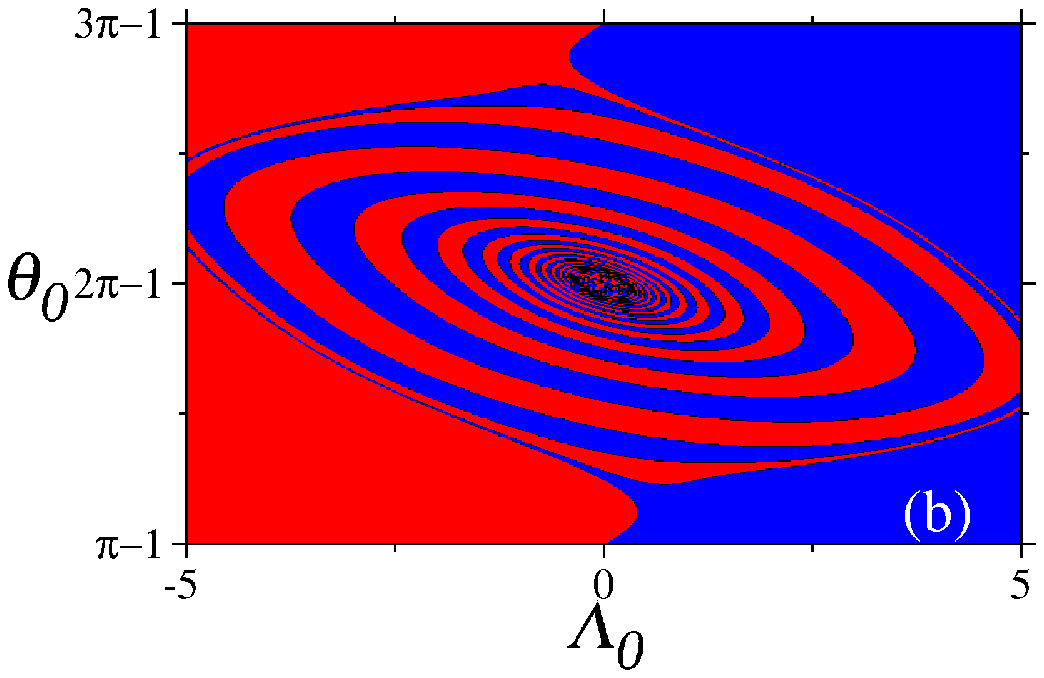}
 \caption{Time averaged angular momentum (color) $\langle \Lambda \rangle$ as a function of the
 ICs ($\Lambda_{\rm 0},\Theta_{\rm 0}$) for $\xi=2.0$, $\gamma=1$, $\eta_0=10$ and $\qA(\tau\!=\!0)=\qB(\tau\!=\!0)=0$. In (a) $\chi=0.7$ and 
$\beta=\pi$. Marked with white lines is the width of several regions with the same $\langle \Lambda \rangle$. In (b) $\chi=\pi /4$ and $\beta=2$ (notice the shifted $\theta_0$-axis). The color coding is as in \fig{220613fig1ab}.}
  \label{220613fig3ab}
\end{figure}

Individual alterations of the parameters $\beta$ or $\chi$ (e.g.\ just $l_{{\rm A}}\neq l_{\rm B}$) do not lead to a preferred rotational direction, since for every solution $[\theta (t),\Lambda (t)]$ of equations \eq{total1} and \eq{total2} there is another solution $(\tilde{\theta}=-\theta (t)+\phi,\tilde{\Lambda}=-\Lambda(t))$, where $\phi$ is a fixed angle. Although those changes do not introduce a preferred rotational direction, it is instructive to understand the effects of changing $\beta$ and $\chi$ in equations \eq{total1} and \eq{total2}. Examples can be seen in \fig{220613fig3ab}, where $\langle\Lambda\rangle$ is plotted as a function of the ICs $(\Lambda_0,\theta_0)$ again. Red stands for clockwise (negative $\Lambda$) and blue for counter-clockwise (positive $\Lambda$) rotational direction. Notice that although $\langle\Lambda\rangle$ changes sign across the \icspace{}, its absolute value is constant. In \fig{220613fig3ab} (a) one sees the case $l_{\rm A}\neq l_{\rm B}$ ($\chi\neq\pi /4$) and $\beta=\pi$, while in b) the case $l_{\rm A}=l_{\rm B}$ ($\chi=\pi / 4$) and $\beta=2$ is shown. The \fig{220613fig3ab} (a) looks qualitatively similar to \fig{220613fig1ab} (a) and (b).
Again, one has fixed points (the same as for the symmetric rotor) and a patterned structure that spirals towards them. Nevertheless, the point anti-symmetry around the fixed points $\theta=\pi /2$ and $\theta=3\pi/2$ ($\Lambda=0$) is no longer present. 
This can be seen by comparing the lengths of the line segments 1 and 2 (3 and 4) in \fig{220613fig3ab} (a), which are the widths of regions with the same $\langle\Lambda\rangle$ measured symmetrically with respect to the aforementioned fixed points.
 The point anti-symmetry around the fixed points $\theta=0$ and $\theta=\pi$ is preserved though.
\Fig{220613fig3ab} (b) presents other differences when compared to \fig{220613fig1ab} (a) and (b): Two fixed points disappear and other two are localized at different positions, namely $\theta=2\pi-\beta / 2$ and $\theta=\pi-\beta / 2$. Nevertheless, the \icspace{} is still point anti-symmetric with respect to the fixed point positions, and therefore again one has $\overline{\langle\Lambda\rangle}=0$. The complete breakage of these point anti-symmetries is just possible by setting both $\chi\neq\pi / 4$ {\it and} $\beta\neq\pi$, and it is exactly this symmetry-breaking that will be exploited to generate directed rotational motion in the next section.

\subsection{Changing both $\beta$ and $\chi$: \adticMotAsym}
In order to obtain a preferred rotational direction one needs to modify $l_{\rm A}\neq l_{\rm B}$ ($\chi\neq\pi /4$) and $\beta\neq\pi$ simultaneously. 
The resulting set of equations does not feature pairs of solutions $(\theta(t),\Lambda(t))$ and $(\theta'(t),\Lambda'(t))$ such that $\theta'(t)=-\theta(t)+\phi$ and $\Lambda'(t)=-\Lambda(t)$. 

To quantify the resulting \adticMotAsym{} we define a \asymMeas{} in the \icspace
\begin{equation}
    M=\frac{\overline{\langle\Lambda\rangle}}{\sqrt{\overline{\langle\Lambda\rangle^2}}}\;,
    \label{asymmetryMeasure}
\end{equation}
where the overbar was defined in equation \eq{overDef}. Note that $M$ is zero when $\overline{\langle\Lambda\rangle}=0$.
For the case where all the $\theta_0$ lead to rotation in the same direction (and with the same angular velocity) one has $M=\pm1$. By changing from 
negative to positive values of $M$ (or vice-versa), we just change
the direction of rotation.

\begin{figure}[tb]
  \centering
  \includegraphics*[width=1\columnwidth]{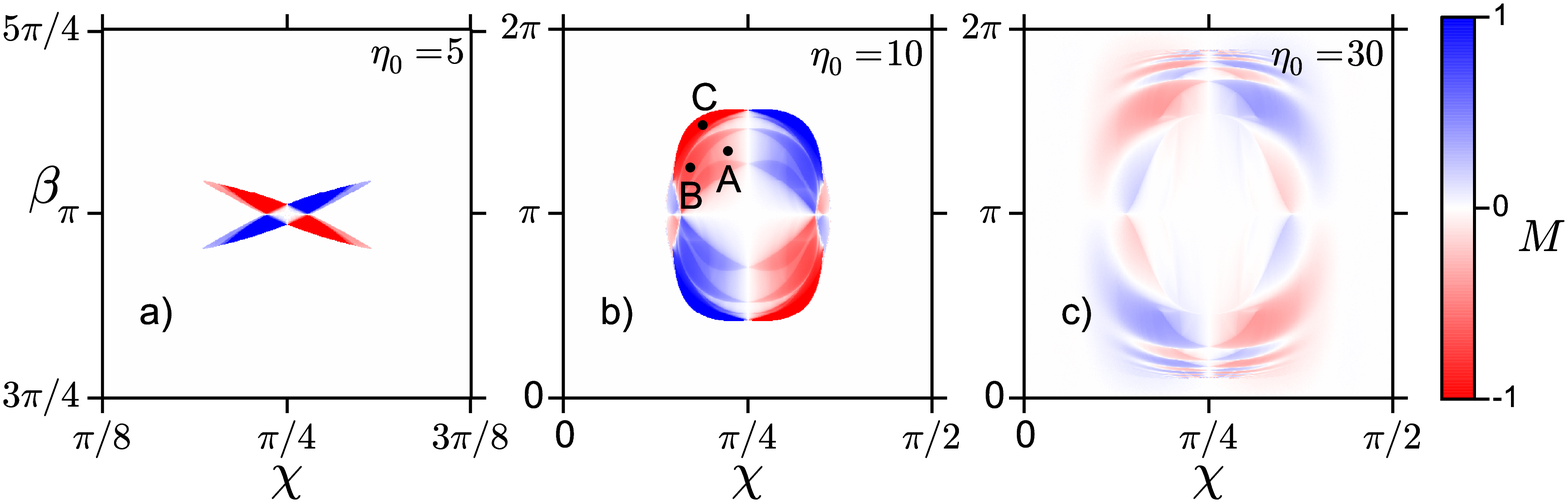}
  \caption{\asymMeas{} $M$ defined in equation \eq{asymmetryMeasure} as a function of the
 parameters $(\chi,\beta)$, for $\xi=2$ and $\gamma=1$.} \label{parspace}
\end{figure}

The \fig{parspace} shows $M$ as a function of the parameters 
$\beta$ and $\chi$. Black indicates no rotational directionality, while from blue(red) to cyan(yellow) an increasing value of the \asymMeas{} is found, with a positive(negative) $\overline{\langle\Lambda\rangle}$. In \fig{parspace} we set the parameter $\gamma=1$ and vary $\eta_0$ (in a) to c)). Along the two lines $\beta=\pi$ and $\chi=\pi/4$ we have $M=0$, as expected. The $M$ values are also \antisymticPS{} about those lines and, consequently, point-symmetric with respect to $(\chi=\pi/4,\beta=\pi)$. This point-symmetry steams from the analytic form in which the parameters $\chi$ and $\beta$ appear into equation \eq{total1}, and therefore it is present independently of the chosen $\gamma$ and $\eta_0$ (and $\xi$). In \fig{parspace} a) ($\eta_0 / \gamma = 5$) and b) ($\eta_0 / \gamma = 10$) there are regions with $|M|=1$, in which a total rotational directionality is achieved. For instance, the point C in \fig{parspace} b) has $M=1$. 
In this case all the ICs lead to the same direction and magnitude of rotation (even for large $|\Lambda_0|$). However, in \fig{parspace} a) almost all the parameter space presents $M=0$, since for such small ratio $\eta_0 / \gamma$ the rotor has $\langle\Lambda\rangle=0$ for most of the $(\chi,\beta)$ values.  Regions where $|M|<1$ are also displayed in \fig{parspace} b) and c) ($\eta_0 / \gamma = 30$). Interestingly, \fig{parspace} c) presents no $|M|=1$ regions, although larger extensions of the parameter space have $|M|>0$. No significant changes in these figures are verified when $\xi=1$ or $\xi=4$. 

\begin{figure}[tb]
  \centering
  \includegraphics*[width=0.49\columnwidth]{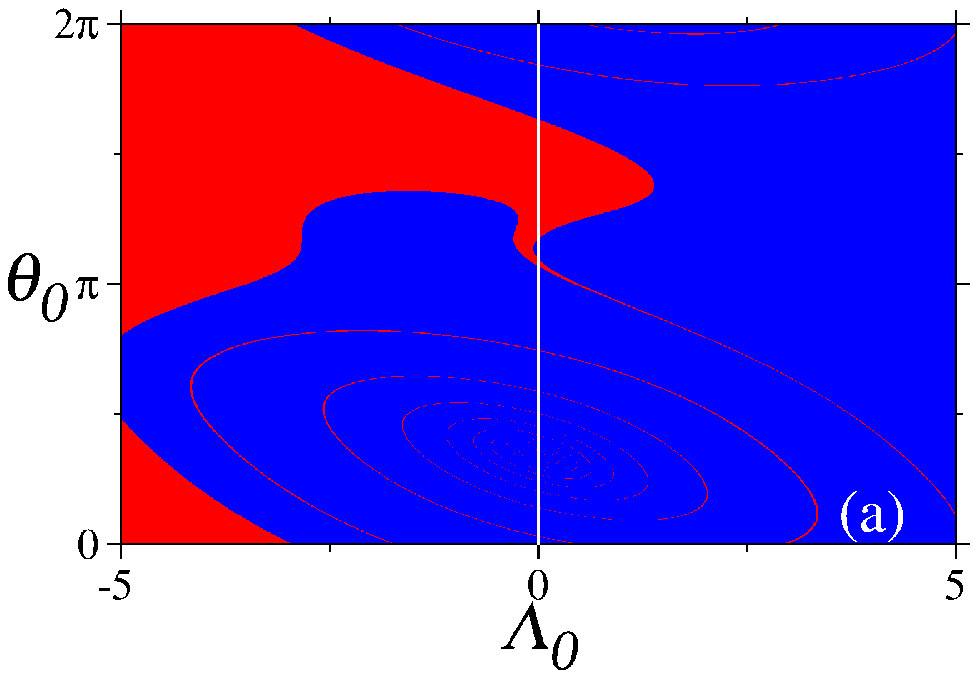}
  \includegraphics*[width=0.49\columnwidth]{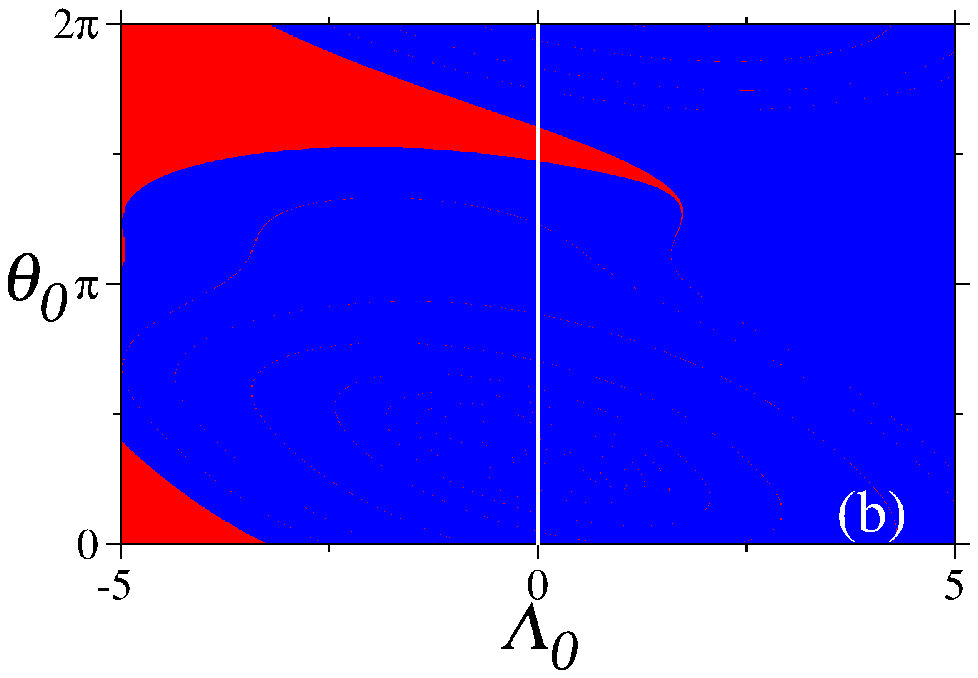}
 \caption{The averaged angular momentum $\langle\Lambda\rangle$ 
as a function of the ICs $(\Lambda_{\rm 0},\theta_{\rm 0})$ for the set of parameters correspondent to a) point A $(\chi=0.7,\beta=4.2)$, b) point B $(\chi=0.54,\beta=3.92)$ in \fig{parspace} b). The color coding is as in \fig{220613fig1ab}. Along the white line in the figures the \asymMeas{} was calculated, as a) $M\approx 0.47$ and b) $M\approx 0.88$.}
  \label{impact}
\end{figure}

Examples for the cases $1>|M|>0$ are the points A and B. For these points we generate the \icspace{} of \fig{impact}. In \fig{impact} (a)((b)) we plot $\langle\Lambda\rangle$  in the \icspace{} for the same parameters as in the point A(B) of \fig{parspace}. There, we find $M=0.47$($M=0.88$). The white line is drawn in \fig{impact} just with the purpose of indicating the locus where the IC-average is performed. By comparing the segments of the white line which are inside regions with positive and negative $\langle\Lambda\rangle$ in both \fig{impact} a) and b), and noting that $|\langle\Lambda\rangle|$ is the same for both red and blue regions, one can understand why the point B presents a stronger rotational directionality than point A in \fig{parspace} b). 
We do not show a figure with $\langle\Lambda\rangle$ as a function of the ICs for the same parameters as on the point
C of \fig{parspace} (b) because {\it all} the ICs lead to the same $\langle\Lambda\rangle$ value (even for $\Lambda_0\neq 0$).

\section{Nanoscale rotor: Stochastic tunneling}
\label{sec:stochastic}
If the size of the rotor (in particular the islands) is shrunk, such that only single electrons are transferred, the mean field
    description used in Sec.\ \ref{sec:model} is not suitable. The charge fluctuations can be of the order of the mean values and it is more appropriate to consider the (dis)charging of the islands as a \stochJ{}.
Now the islands can either be empty or occupied by one electron, so that the charges $Q_{\rm A}$ and $Q_{\rm B}$ can be either $0$ or $-e$. 
 The occupation in 
the islands changes from $0$ to $-e$ or vice versa whenever a tunneling event occurs. The tunneling 
is modeled as a stochastic event.
The probability  for a tunneling event during  an infinitesimal time 
interval $d\tau$ is
\numparts
\begin{eqnarray}
&{\rm d}P_{{\rm in},i}={\rm d}\tau W_{\rm in}(\theta'), \\
&{\rm d}P_{{\rm out},i}={\rm d}\tau W_{\rm out}(\theta'),
\end{eqnarray}
\endnumparts
where $dP_{{\rm in},i}$($dP_{{\rm out},i}$) is the infinitesimal probability of tunneling in(out) the island $i=A,B$, and $\theta'=\theta$ for $i=A$ and $\theta'=\theta+\beta$ for $i=B$.

In order to assess the impact of this stochastic (dis)charging on the \adticMotAsym{} we exemplarily calculate \asymMeas{} as a function of the parameters $(\chi,\beta)$. Since the tunneling is a stochastic process, one needs to redefine the \asymMeas{}. Instead of considering $\langle\Lambda\rangle$ and $\langle\Lambda\rangle^2$, we consider their average over the realizations $\langle\langle\Lambda\rangle\rangle_{\rm re}$ and $\langle\langle\Lambda\rangle\rangle_{\rm re}^2$, such that the new \asymMeas{} is
\begin{equation}
    M_{\rm stoch}=\frac{\overline{\langle\langle\Lambda\rangle\rangle_{\rm re}}}{\sqrt{\overline{\langle\langle\Lambda\rangle\rangle_{\rm re}^2}}}\;,
    \label{aMeasStoch}
\end{equation}
where $\langle ...\rangle_{\rm re}$ is an average over realizations and the overbar the same average over ICs as in equation \eq{asymmetryMeasure}. The \asymMeas{} $M_{\rm stoch}=0$ when $\overline{\langle\langle\Lambda\rangle\rangle_{\rm re}}=0$. For $M_{\rm stoch}=\pm 1$ one has {\it all} ICs leading to net rotations in the same direction with the same $\langle\langle\Lambda\rangle\rangle_{\rm re}$, such that the sign just indicates the direction of rotation. It must be noted that the sign of $\Lambda(t)$ within a {\it single trajectory} can change with time, even though we have $M_{\rm stoch}=\pm 1$.

\begin{figure}[tb]
  \centering
  \includegraphics*[width=1\columnwidth]{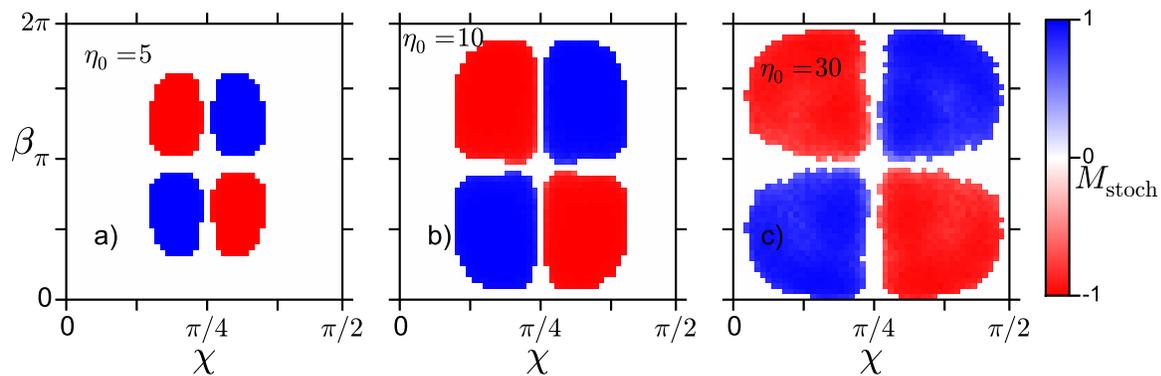}
  \caption{Directionality measure $M_{\rm stoch}$ defined in equation \eq{aMeasStoch} as a function of the
 parameters $(\chi,\beta)$, for $\xi=2$ and $\gamma=1$ (same parameter set and color coding as in \fig{parspace}).}
  \label{stoch}
\end{figure}

The \fig{stoch} shows $M_{\rm stoch}$ as a function of the parameters 
$\beta$ and $\chi$ for the same $\gamma$, $\eta_0$ and $\xi$ as in \fig{parspace}. Black indicates no rotational directionality, while from blue(red) to cyan(yellow) an increasing value of the \asymMeas{} is found, with a positive(negative) $\langle\Lambda\rangle$. $M_{\rm stoch}$ presents the same symmetry in the parameter space as $M$ does. Along the two lines $\beta=\pi$ and $\chi=\pi/4$ we have  the expected $M_{\rm stoch}=0$ value. The $M_{\rm stoch}$ values are also \antisymticPS{} about those lines and, consequently, point-symmetric with respect to $(\chi=\pi/4,\beta=\pi)$. Again, this symmetry is independent of the chosen $\eta_0$ and $\gamma$. The \fig{stoch} features extensive regions of total or partial rotational directionality, namely in the red-yellow and blue-cyan areas. This represents a strinking difference to the mean-field equations results from \fig{parspace}. The results also do not depend on $\eta_0 / \gamma$ so sensitively as in the mean-field case (compare \fig{stoch} with \fig{parspace}). Nevertheless, the $|\langle\langle\Lambda\rangle\rangle_{\rm re}|$ achieved are smaller than the typical $\langle\Lambda\rangle$ from the mean-field case. This can be partially explained by the fact that $\Lambda(t)$ can change its sign within a single trajectory.

\section{Conclusions}
\label{sec:conclusion}
In the present work we have studied how to control the rotational direction of an electromechanical rotor. We describe the charge dynamics in the islands using two different approaches: a mean-field approach, where charge fluctuations are disregarded; and a \stochJ{} approach, where the charge transfer occurs via stochastic tunneling events. The first approach is valid for a macroscopic electromechanical rotor (see appendix \ref{sec:rotor_setup_class}), while the second one is more appropriate for a nanoelectromechanical device where the Coulomb blockade effect is relevant.

In the mean-field approach, rotational direction control can be achieved in two different ways: one can either select the rotational direction via choosing an appropriated IC, or one can break the symmetry of the rotor to introduce a preferred rotational direction. In order to experimentally perform the first method in a macroscopic system, the precision of the IC preparation must be high enough to resolve structures in the \icspace{}. While our mean-field calculations show that increasing the ratio $\eta_0/\gamma$ also increases the \icspace{} size of the structures, through the \stochJ{} calculations one sees that charge fluctuations can destroy the dependence of the system on the IC. This means that an IC preparation scheme may fail for a real nanoelectromechanical rotor. Apart from that, fabrication imperfections can also spoil the attempts of selecting a specific IC which leads to a certain rotational direction. In contrast, our numeric simulations show that, by breaking the rotor's symmetry, most or even {\it all} ICs lead to the same rotational direction. For that reason, we focused on the symmetry-breaking procedure.

We have shown that one needs to change the system's parameters $\chi$ and $\beta$ {\it simultaneously} in order to break the symmetry of the rotor. The parameter $\beta$ is tuned by setting a tilting angle between the two rotor's arms. We have considered the ratio between the arms' lengths to determine $\chi$, but another way would be having islands with different maximal charges $Q_{\rm maxA/B}$. The parameter space of both mean-field and stochastic equations features extensive areas where a total rotational directionality (with \asymMeas{} modulo 1) is present. In these areas, \icspace{} scans indicate that the same $\langle\Lambda\rangle$($\langle\langle\Lambda\rangle\rangle_{\rm re}$) in the mean-field(\stochJ{}) approach is obtained {\it independently} of the chosen IC. The symmetry-breaking procedure works for large sections of the parameter space. In an experimental realization of our equations, imperfections in the rotor fabrication within the limits of these sections would not spoil the symmetry-breaking. Additionally, our stochastic simulations feature large regions in the parameter space leading to rotational directionality. Thus, this procedure may work specially well for nanoelectromechanical rotors.

\ack
MWB thanks CNPq (Brazil) for financial support.

\appendix

\section{Derivation of the mechanical equations of motion}\label{derivAng}

In this section we derive equations \eq{total1} and \eq{total2}, which determine the dynamics of the rotor's mechanical degrees of freedom. The torque 
$T$ that acts on the rotor is given by
\begin{equation}
T=\frac{\partial}{\partial t}L=l_AEQ_A\sin{\theta}+l_BEQ_B\sin(\theta+\beta)-\lambda L,
\label{torque}
\end{equation}
where $L$ is the angular momentum of the rotor. The equation for the angle $\theta$ is
\begin{equation}
\frac{\partial}{\partial t}I\theta=L,
\label{theta}
\end{equation}
where $I$ is the moment of inertia of the rotor. Dividing equation \eq{torque} by $Q_{\rm max}I/t_0^2$ one obtains
\begin{eqnarray}
\label{willlead}
\fl \frac{\partial}{\partial t/t_0}\left(\frac{L t_0/I}{Q_{\rm max}}\right)&=\frac{l_AE}{I/t_0^2}\frac{Q_A}{Q_{\rm max}}\sin{\theta}+\frac{l_BE}{I/t_0^2}\frac{Q_B}{Q_{\rm max}}\sin(\theta+\beta)-\frac{\lambda}{I/t_0^2Q_{\rm max}} L.
\end{eqnarray}
Using dimensionless parameters as defined in Table \ref{dlqtts} and $\gamma=\lambda t_0$ one obtains
\begin{equation}
\frac{\partial}{\partial \tau}\Lambda = -\eta_0\sqrt{2}\left(\sin{\chi}\sin{(\theta)}\qA+\cos{\chi}\sin{(\theta+\beta)}\qB
    \right)- \gamma\Lambda,
\end{equation}
which is identical to equation \eq{total1}.

Equation (\ref{total2}) is obtained after dividing Eq.~\eq{theta} by $I/t_0$.

\section{Macroscopic realization of the mean field equations}\label{sec:rotor_setup_class}
According to equations \eq{eq:dtQ1} and \eq{eq:dtQ2}, the charge dynamics inside an island is determined by
\begin{equation}
\dot{Q}=W_{\rm in}(\theta')(Q_{max}-Q)-W_{\rm out}(\theta')Q,
\label{objective}
\end{equation}
where $Q$ is the charge on the island with maximum value $Q_{\rm max}$ and $W_{\rm in/out}$ are periodic functions of
$\theta'$ of the form
\begin{eqnarray}
W_{\rm in}(\theta')&=\frac{1}{t_0}{\rm e}^{-\xi \cos{\theta'}},\\
W_{\rm out}(\theta')&=\frac{1}{t_0}{\rm e}^{\xi \cos{\theta'}}.
\end{eqnarray}
One could think of realizing Eq.~(\ref{objective}) 
by using a RC circuit with two capacitors with capacitances $C_{\rm in}$ and $C_{\rm out}$ configured in series
\begin{equation}
V=R\dot{Q}+Q\left (\frac{1}{C_{\rm in}}+\frac{1}{C_{\rm out}}\right )=R\dot{Q}+Q\left (\frac{C_{\rm in}+C_{\rm out}}{C_{\rm in}C_{\rm out}}\right ),
\label{compare}
\end{equation}
where $V$ is the voltage of the circuit and $C=(C_{\rm in}+C_{\rm out})/(C_{\rm in}C_{\rm out})$ is the effective capacitance.
Comparing Eqs.~(\ref{objective}) and (\ref{compare}) one obtains
\begin{eqnarray}
W_{\rm in}(\theta')&=\frac{1}{C_{\rm in}(\theta')R},\\
W_{\rm out}(\theta')&=\frac{1}{C_{\rm out}(\theta')R},
\end{eqnarray}
or
\begin{eqnarray}
C_{\rm in}(\theta')&=\frac{t_0}{R}{\rm e}^{-\xi \cos{\theta'}},\\
C_{\rm out}(\theta')&=\frac{t_0}{R}{\rm e}^{\xi \cos{\theta'}}.
\end{eqnarray}
Finally, the voltage also has an angular dependence, namely
\begin{equation}
V(\theta')=\frac{Q_{\rm max}}{C_{\rm in}(\theta')}.
\end{equation}

\bibliography{arxiv_32}

\end{document}